\documentclass[manuscript]{aastex}

\shorttitle{Beyond Nadel's Paradox. A  computational approach to structural and cultural dimensions of social cohesion} \shortauthors{Lozano et al.}

\begin{document}

\title{Beyond Nadel's Paradox. A  computational approach to structural and cultural dimensions of social cohesion}

\author{S. Lozano}
\affil{ETH Zurich, Swiss Federal Institute of Technology, Zurich, Switzerland.}

\author{J. Borge}
\affil{Department of Psychology, Universitat Rovira i Virgili,
Tarragona, Spain}

\author{A. Arenas}
\affil{Department of Computer Sciences and Math, Universitat Rovira
i Virgili, Tarragona, Spain}

\and

\author{J.L. Molina}
\affil{Department of Social and Cultural Anthropology, Universitat Autonoma de Barcelona, Barcelona, Spain.}

\begin{abstract}
Nadel's Paradox states that it is not possible to take into account simultaneously cultural and relational dimensions of social structure. By means of a simple computational model, the authors explore a dynamic perspective of the concept of social cohesion that enables the integration of both structural and cultural dimensions in the same analysis. The design of the model reproduces a causal path from the level of conflict suffered by a population to variations on its social cohesiveness, observed both from a structural and cognitive viewpoint. Submitted to sudden variations on its environmental conflict level, the model is able to reproduce certain characteristics previously observed in real populations under situations of emergency or crisis. \\
\end{abstract}

\keywords{social cohesion, dynamic analysis, social structure, conflict, social networks}

\section{Introduction}
Paul Dimaggio \citep{Dimaggio} remembers us the Nadel's Paradox, who followed the distinction done for Radcliffe-Brown \citep{Radcliffe-Brown} and the British structural-functionalist school between culture and structure:

\textit{This is Nadel's Paradox: A satisfactory approach to social structure requires simultaneous attention to both cultural and relational aspects of role-related behavior. Yet cultural aspects are qualitative and particular, pushing researchers toward taxonomic specifity, whereas concrete social relations lend themselves to analysis by formal and highly abstract methods.} (\citep{Dimaggio}:119-120).

Nadel \citep{Nadel} was interested in describing social structure as a web of roles, which were composed both by a pattern of relationships with other roles as by a cultural content observable in behavior. While getting knowledge of the pattern of relationships implied formal operations and allowed consequently a strong operacionalization, the description of variability of role's cultural content leaded to broad typologies and implied a weak operacionalization.  At the same time, the structural patterns could be aggregated in greater structures while cultural based typologies only could be considered substantively. This weak operacionalization is common in Social Sciences: 'Mind' in the psychological literature, 'culture' in anthropology or 'wealth' in economics, for instance, they all share a vague definition, due both to the central role they play in their disciplines, and to a long and abundant literary history.

This paradox is real, and it represents the different traditions in Social Sciences and Humanities among other dualities, as the pairs qualitative-quantitative and structure-agency, for instance. From our point of view, social networks perspective allows overcoming some of those dualities for its capacity for being situated among the \emph{macro}, historical and institutional level, and the \emph{micro}, the individual and biographic agency, the \emph{mesolevel}. Moreover, a \emph{dynamic} perspective of social networks allows taking into account social norms culturally determined \emph{and} social patterns of relationship among individuals and groups. For illustrating this point we will concentrated in \emph{social cohesion}, especially for the role played for it in the emergency of social movements in moments of crisis.

Social cohesion is also a concept with a long history and multiple definitions \citep{Kearns,Friedkin}. Needless to say, social cohesion has been an issue under study in sociology since its modern foundations \citep{Durkheim} , affecting many related issues such as community structure \citep {Forrest} or population health \citep{Wilkinson, Kawachi} and, as a consequence, an attempt to develop a theory of social cohesion is confronted with partial definitions of it, complex literature from different sociological schools or unsuited approaches \citep{Friedkin}. From a theoretical point of view, for instance, the concept is diffuse because of the difficulties in isolating its meaning, which overlaps several other core concepts such as solidarity, inclusion or integration. As an example, Kearns and Forrest \citep{Kearns} distinguish as much as five different facets of social cohesion (which they name \emph{constituent dimensions}), including common culture, solidarity, networks and social capital, and territorial belonging and identity. These manifold features entail intrinsic methodological difficulties, since they include cognitive, social, structural and spatial dimensions. Moreover, from the methodological standpoint it is not clear whether social cohesion is to be understood as a cause (or independent variable) of other phenomena, such as economic performance \citep{Wolfe} or well-being \citep{Wilkinson}, or as a consequence (dependent variable) from other phenomena, i.e. community dynamics \citep{Putnam} or economic restructuring \citep{Polanyi}.\\

The need of a certain degree of consensus around the definition of cohesiveness is central to Social Sciences, since it is related to many phenomena of interest, such as community conceptualization \citep{Wellman&Leighton, Wellman&altri}, social exclusion and integration \citep{Room} or the concept of embeddedness \citep{Granovetter,Granovetter2}, mainly relevant in economic sociology.\\

Returning to the first paragraphs, a common strategy to short-circuit such millstones in social sciences is operationalization, i.e. defining methodological and quantification paths towards the fixation of the concept. For instance, such scheme lies at the kernel of cognitive psychology, which has turned the ethereal cartesian 'mind' into an information-processing construct.\\

Beyond its theoretical conceptualization, attempts to operationalize cohesion have appeared, mainly from a social networks approach \citep{Wasserman,Moody&White}, which is specially well-suited for the mathematical treatmement of social interaction. One approach in this line which has received a significative attention is \emph{Group structural cohesion}. It applies both to small sets and larger scales, and is defined as \emph{"the minimum number of actors who, if removed from the group, would disconnect the group"} \citep{Moody&White}. Notice that, as the authors themselves highlight, such definition's scope is limited to relational (node connectivity) aspects, and fails to capture other relevant dimensions of cohesion and related concepts like \emph{solidarity} and \emph{embeddedness}.\\

Thus, operationalization is not a final solution, but a means to re-locate and focus the study of the fundamental issues. Accordingly, the aim
of this paper is to broaden Moody and White's influential proposal
on structural cohesion, by integrating a cognitive, cultural component to the
already-existing structural one in dynamical environments.

In particular, we will argue that social cohesion is a process through time, genuinely a dynamical concept. Given a network topology at a moment in time, the measure of structural cohesion fails to grasp other dimensions of the concept, while an evolutive treatment of it allows the inclusion of the cultural and cognitive elements, which may be stated as follows \citep{Carley}: (1) Individuals are continuously engaged in acquiring and communicating information; (2) what individuals know influences their choices of interaction partners; and (3) an individual's behavior is a function of his or her current knowledge.\\

To make such an evolutive treatment, we have developed a computational model. As far as we know, previous works addressing this topic by means of this sort of tools, have fulfilled the requirements above only partially. One significant example is \citep{Vega-Redondo}, where authors analyze the structural cohesion of a population of individuals (in terms of its network's resilience to a dynamic, uncertain, scenario). The computational model presented there, captures the interplay between information and network dynamics (first two points in the list above), but does not include the cultural component we are stressing here (third requirement).        

In the remainder of this introduction, we give a brief discussion of the main components of our hypothesis, which include a short comment on the cohesion concept based on the analysis of three bibliographic examples, and some sociological background considered for the design of the model presented in the paper. The second subsection is devoted to a detailed description of the model's main components. Finally, our results are summarized in the third part, along with evidence from historical sociology that provide empirical support to our claims.\\

\subsection{Empirical motivation}
In order to outline the purpose of this work, we offer three examples taken from  sociology \citep{Gould}, economics \citep{Stark} and anthropology \citep{Murphy} that may illustrate it.\\

Gould's paper \citep{Gould} analyzes insurgent activity during the Paris Commune in 1871, which sprung after a mixture of political, economic and war crises. In this paper, as in later works \citep{Gould_llibre, Gould_paper} Gould settles, by means of data analysis, that organizational networks and pre-existing informal networks interacted in the mobilization process. As Gould points out, mobilization does not just depend on existing social ties; it also creates them. Although members of a protest organization may have joined because of a pre-existing social tie to an activist, they also form new social relations while participating in collective protest.\\

Stark's work \citep{Stark} on economic dynamics covers a time interval centered on the transition years (from a communist regime towards an open market economy) in Hungary, which were characterized by political and economic discontentment. Stark analyzes cohesive processes at the economic level by means of large data sets, studying the formation (and dissolution) of clusters of firms. Examining Hungary's political history in that period of time and Stark's economic survey, it is possible to observe a certain correlation between both processes. In the early 90's, economic cohesion appears to evolve as a delayed result of civil unrest. Stark's work will be more thoroughly developed in section 3.\\

Finally, warfare patterns are the main concern in Murphy's study
\citep{Murphy} of a Brazilian Indian group, the Munduruc\'{u}. Mostly in the 19th century, the Munduruc\'{u} participated in many raids, mainly for cultural reasons, secondary as a mercenary service for the Brazilians. Besides its anthropological interest, the remarkable fact is the organizational aspects of warfare within this Amazonian culture. The Munduruc\'{u}, although a cultural unity, was settled in several apart villages, spread along the upper Tapajós River. However, the setup of war parties evidenced a strong relationship among these communities, otherwise unobservable: intercommunity cooperation in warfare was facilitated by cross-cutting ties of residential affinity and affiliation by descent. For example, any Munduruc\'{u} man was a member of his own village, in most cases a native of another village, and in several instances a former resident of the village of a divorced wife; he was also linked by ties of patrilineal descent to all the villages in which members of his clan resided. Therefore, he was not involved in any cohesive and localized lineage unit. The kinship structure imposed no strict boundaries upon the local group and the male social world was widely distributed.\\

After these short resumes, we can point out some remarkable facts
observed in the three cases despite corresponding to different
scenarios and scholar approaches:
\begin{itemize}
\item All the approaches here presented are focused on \emph{processes}, rather than static \emph{situations}, they consider different kind of events through time.
\item Some type of conflict is present and central to each work. Notice that \emph{conflict} is here understood in a broad manner, and therefore it might sometimes imply violence, sometimes passive resistance, etc.
\item The structural characteristics of the system after the conflict occurs is different from the structural characteristics before it. Some of the work presented settles this fact explicitly, like Gould's, while the data presented in others, like Stark's, state it implicitly.
\end{itemize}

Underlying all these formal resemblances rest the main ideas that
constitute our proposal's framework. Those ideas, that we have tried
to capture in our model's design, are listed in the following.\\

Taking structural and cognitive components of cohesion into account is (only) possible through dynamic observation of a system. It can be said that while structural cohesion is observable in static environments, cohesion growth depends on mobilization and recruitment that take place through \emph{changes} at the individual level \citep{Snow}.\\

Explaining dynamical behavior demands to elucidate how macro, meso and micro variables interact during conflict, and how they affect cohesion. Social movement behaves in a regular pattern; from the institutional (macro) level to the cognitive, psychologic (micro) sphere through the (meso) level of networks and vice versa \citep{Coleman}. This interaction at the meso level is complex and it is constituted both for processes of selection on the part of individual and influence by groups \citep{Snijders&altri}.\\ 

In other words, a mobilization begins with a mobilization potential which depends both on macrostructural factors such as demographic, economic or ideological variables and individuals predispositions and social networks structures in which they are embedded, who, in turn, change their connectivity thus affecting social groups' structure and the macrostructural framework.\\

It is possible, considering the previous items, to observe
\emph{covert cohesion}, i.e. beyond existing ties among agents,
conflict activates potential, previously nonexistent links, which
effectively cause cohesion growth. Conflict itself is not a direct
cause of the observable changes in structural cohesion, but a link
in a causal chain; instead, clash affects information exchange among
agents (consciousness-raising, \citep{Hirsch}), triggers possible
out-group ties between subjects and by doing so uncovers an existing
cohesion which depends both on structure and cognitive level. Covert
cohesion, then, takes it form at the cognitive level, and is expressed
at the group level by means of increasing structural cohesion.\\

The computational model described in the next subsection assumes the mentioned causal relationship, therefore bestowing structural cohesion with the cognitive component, being then an appropriate implementation of the described hypothesis.

\subsection{Modelling of Social Cohesion: Backgrounds}
The preceding examples illustrate that there are underlying mechanisms which allow the emergence of cohesion. However, such approach is mainly qualitative, whereas this work is concerned about simulation and quantification of such underlying mechanisms. The model we introduce is meant to implement plausible dynamic behavior, attending the mainstream sociological theories on interaction among agents.\\

Although there is a growing literature in dynamic models of networks analysis \citep{Wasserman&Robins, Snijders, Vega-Redondo} and other for simplification purposes, the description of the model will follow Carley's discussion about group stability. Following Carley's theory \citep{Carley}, realistic social modelling ought to incorporate (1) a structured model of information, (2) a model of information forgetting, (3) institutional or environmental limits on interaction or forced interaction, (4) a model of population dynamics, or (5) a model of information discovery.\\

Regarding social dynamics (4 in Carley's list), we have followed Lazer's directions on coevolutive systems \citep{Lazer}. According to him, social change is not to be understood \emph{solely} in terms of the context, nor \emph{solely} on individual choices. On the contrary, an appropiate account for social dynamics is to be found somewhere within those two extremes: individuals' choices are molded by the network, and individuals' behaviour actively affects the network as a whole. In order to provide for such requirement, the model's dynamic is driven by a co-evolution model proposed by Holme and Newman in  \citep{Holme&Newman}, which assumes Lazer's premises on coevolution, i.e. the system evolves in a twofold way: individuals become likeminded because they are connected via the network (change is induced by the structure) or they form network connections because they are like-minded (structure undergoes change). It does so by either redefining the connectivity of the population of agents or by changing their positioning in the social space.\\

However, since one of the major elements of integration is the extent to which various members interact with one another, modeling social cohesion demands not only a dynamical characterization, but also needs to include the notion of social distance (1 in Carley's list). Following Blau \citep{Blau,Blau2} and Fararo and Skvoretz \citep{Fararo} we can fulfill such need. According to them, interaction among agents depends on the number of dimensions that people have in common. Unifying Granovetter \citep{Granovetter3}, Blau, Fararo and Skvoretz link interaction and similarity directly on social dimensions, positing that social associations are more prevalent between persons in proximate than those in distant social positions. Such a principle can be translated in probabilistic terms, and thus become part of our computational model.\\

Accordingly, along with dynamic behaviour stated above, the system
comprises social distance treatment. To do so, the model includes a
definition of the linkage probability between two individuals,
proposed in \citep{Boguna&altri}, that
is based on the social distance between them in a social space of a certain dimension. As in Fararo and Skvoretz, this specification weights all dimensions equally.\\

As for item 3 in Carley's list, the equation defining linkage probability includes a parameter that expresses the degree of conflict. This parameter is named \emph{social temperature}, since it allows the model to simulate situations of political disorder, warfare or large demographic changes (high social temperature).\\

There is still another important feature of the model that needs some development, which corresponds to item 2. In the search for a realistic treatment of information exchange among subjects, the system includes a external universal parameter (i.e. it applies to all agents and is independent of the system state), which deviates or modifes their current knowledge in a different way for each member of the population. As it is explained below, we name this parameter \emph{social noise}.\\

\section{COHESION ANALYSIS THROUGH A COEVOLUTIVE MODEL}

In the previous chapter, we have developed the main guidelines of our model's design from the analysis of three works in the literature related, in some sense, to social cohesion. In this chapter, we make a complete description of the model, paying especial attention to concepts introduced above such as \emph{social temperature} and \emph{social noise}.\\

\subsection{Complete description of the model}
We consider a population of $N$ agents, connected through a variable
number of undirected (bidirectional) links. Each agent $i$ presents
a $h_i$ value, corresponding to his location in a continuous lineal
social space of size $L$ (proportional to $N$). Taking into account
the explanations of the previous chapter, $h_i$ could be seen as an
opinion or positioning of individual $i$ in relation to a certain topic (related to religion or politics, for instance).\\

Initially the $h$ values of all agents are assigned randomly,
following a uniform distribution along the lineal social space.
Besides, the initial arrangement of the edges correspond to a
topology with the same structural properties than real social
networks (like large clustering coefficient and positive degree
correlations, for instance). To construct such a scenario, we use a
class of models proposed in \citep{Boguna&altri}, that are able to
grow up networks with social-like macroscopical (global) properties
from a microscopical (individual) definition of the linkage
probability between two agents. The key element of that definition,
is the social distance between the two individuals in a social space
of a certain dimension $d_\mathcal{H}\geq 1$. Since our social space
is lineal, here we use a simplified expression of the linkage
probability with $d_\mathcal{H}= 1$:

\begin{equation}
  r(h_i, h_j)= \frac{1}{1+\left[b^{-1} d (h_i,h_j)\right]^{\alpha}}
  \label{rtotal}
\end{equation}

Where $d(h_i,h_j)$ is the social distance, $b$ a parameter
controlling the length scale of the lineal social space, and
$\alpha$ quantifies the homophily, that is, the level of
restrictiveness of an individual to interact with others in function
of their social affinity \citep{McPherson}. So, given a certain
social distance between two agents, different combinations of $b$
and $\alpha$ values lead to different link probabilities, in such a
way that the higher the $b$ and the lower the homophily, the larger the probability of connection.\\

As said in the previous chapter, the model evolves from the initial
scenario in a twofold way, by redefining both the topology of the
network and the positioning of the population of agents in the
social space. Based on a co-evolution model proposed by Holme and
Newman in \citep{Holme&Newman}, the two main mechanisms that boost
this co-evolution process are the rewiring of links and the
imitation of $h$ values among agents. Additionally, we have incorporated a third mechanism that reproduces those little shifts on everyone's opinion or social position, induced by individual circumstances and daily life experiences, that usually modify individuals' knowledge in a subtle but continuous way. This third mechanism is necesarily external, since these particular characteristics are different for each individual, and don't depend on any other parameter of the model. Notice that, at the mid-long time range, these slight but continuous shifts can change significantly the social distance among two individuals,
separating two agents that were once very close in the lineal social
space or, on the contrary, approximating them enough to favor the
creation of a new link. Consequently, the accumulation of these
microscopical changes can modify the whole macroscopical scenario,
by disrupting both the distribution of agents' positions along the
social space and their connectivity. Taking into account this
disrupting effect over the whole system, closely similar to the
concept of noise in physics and electronics, we have denoted this
third mechanism of the dynamics as \emph{social noise}.\\

These three mechanisms (rewiring, imitation and social noise) are
integrated within the co-evolutive dynamics of the model, consisting
on the repetition of the following two steps:

\begin{enumerate}
    \item Select an agent $x$ at random and decide, with equal probability, whether to apply rewiring or
imitation.
    \begin{itemize}
    \item The rewiring consists on a redefinition of all links of
node $x$ using the expression in (1).
    \item Imitation is implemented by selecting randomly a neighbor $y$ of node $x$, and setting $h_y$ equal to $h_x$.
    \end{itemize}

    \item Introduce the \emph{social noise} by summing up a random quantity (obtained from a
    gaussian distribution with a 0 mean and fixed variance) to the $h$ values
    of all agents in the population.
\end{enumerate}

\begin{figure}[b]
\plotone{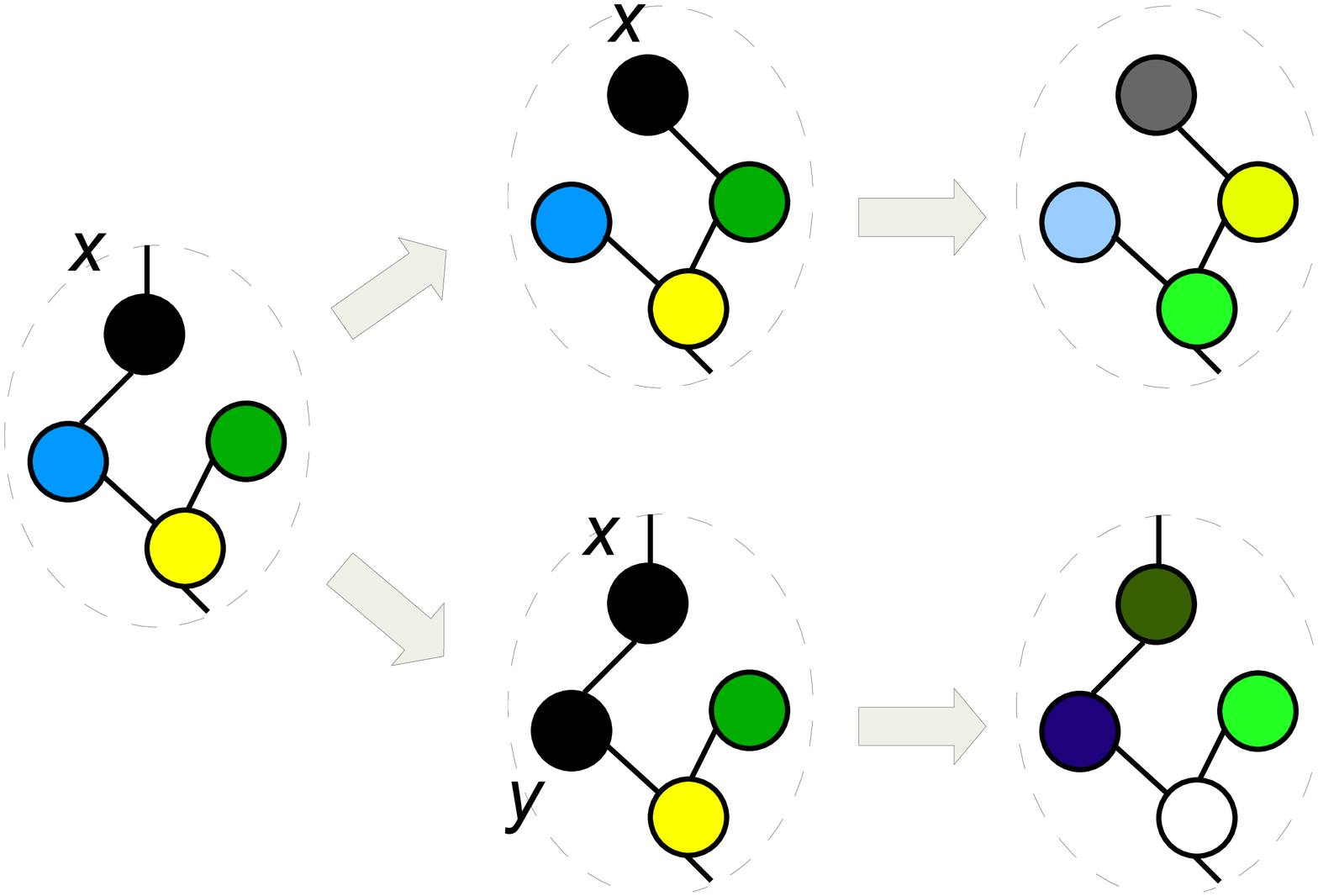}
 \caption{An illustration of our co-evolutive dynamics, where colors indicate the $h$ value of each individual. At each time step, the system evolves following one of the two branches.
  The upper branch correspond to a rewiring (of $x's$ links) plus a shift of all positions, and the lower one to imitation ($y$ imitates $x$) plus position shifts.}
  \label{Fig_dyn}
\end{figure}

Fig. \ref{Fig_dyn} illustrates this dynamics. At each time step, the
system evolves following one of the two possible branches of the
diagram (imitation plus \emph{social noise}, or rewiring plus
\emph{social noise}) with the same probability.\\

After a certain number of time steps, the system reaches a
\emph{steady-state}. In our context, this means that both the
topology and the distribution of individuals' social positions along
the space remain stable. The concrete topology and distribution of
social positions reached in each possible steady-state depend, as we will
show in the next subsection, on the strength of the \emph{social noise}.\\

Finally, two additional elements need to be included in our model to
study social cohesion and its interplay with extremal changes on the
social environment. On one side, it should be able to simulate
different social temperatures (as stated in the previous
subsection). On the other hand, it has to include an observable to
signal how these changes influence the social cohesiveness of the population.\\

The first requirement, introduction of changes on the social
temperature, has been modeled as variations on the value of the $b$
parameter (the one controlling the length scale of the social
space). This solution can be justified as follows. When some kind of emergency strikes a population, social distances that separate individuals do not change, but the necessity to face the new critical scenario makes them less important than in a quiet situation. This temporal relativization of social distances is nothing else than a change on the length of the scale they are 'measured' with. Consequently, an appropriate way to introduce in our model the effect of emergencies and posterior relaxations of the conflict level, is to increase the value of b (making distances relatively smaller) and, after a relatively short number of time steps, reduce it back. $b$ (making distances
relatively smaller) and, after a relatively short number of time steps, reduce it back.\\

Regarding the monitoring of social cohesiveness, we have defined
three different macroscopical observables, namely: the average
degree $\langle k \rangle$ (average number of neighbors), the
clustering coefficient (a weighted measurement of the number of
triangles) and the number of disconnected components or independent
groups $G$ composing the whole network. While first and second parameters signal intra-group cohesion, the third one corresponds to inter-group cohesiveness. Note that, taken jointly, these three are good indicators of the social cohesiveness, since the more cohesive is a population, the higher are their average degree and clustering coefficients, and fewer separate groups it presents.\\

\subsection{Effect of \emph{social noise}}

An important issue to deep in at this point, is the influence of the \emph{social noise} over the evolution of the model. As said before, the strength of the social noise can determine the steady-state, that is, the stable configuration where the co-evolutive model ends up. Since this noise is defined by a gaussian distribution with a fixed mean and variance, we center our attention on the unique parameter of the \emph{social noise} that can be tuned: its magnitude.\\

In the context of our model, the magnitude corresponds to the
average celerity of changes experimented by individuals' knowledge
due to the \emph{social noise}. Large amounts of \emph{social noise}
imply sudden changes of individual's social positions along the
social space. On the contrary, low noises correspond to quite stable opinions.\\

Taking this into account, we can easily predict the behavior of the
model for extremal values of the noise magnitude. On one side, too
much \emph{social noise} would result in a noise-dominated scenario,
where agents would be almost completely isolated due to the
difficulty to maintain links among them. On the other hand, if the
noise was too less intensive it would exercise no significant effect
over the dynamics, which would be controlled by the other two
mechanisms (imitation and rewiring). Keeping this in mind, some
questions arise: what are we to understand as "too weak" or "too
strong" noise? And, how does the noise influence the
dynamics for intermediate strength values between these limits?\\

\begin{figure}[t]
\plotone{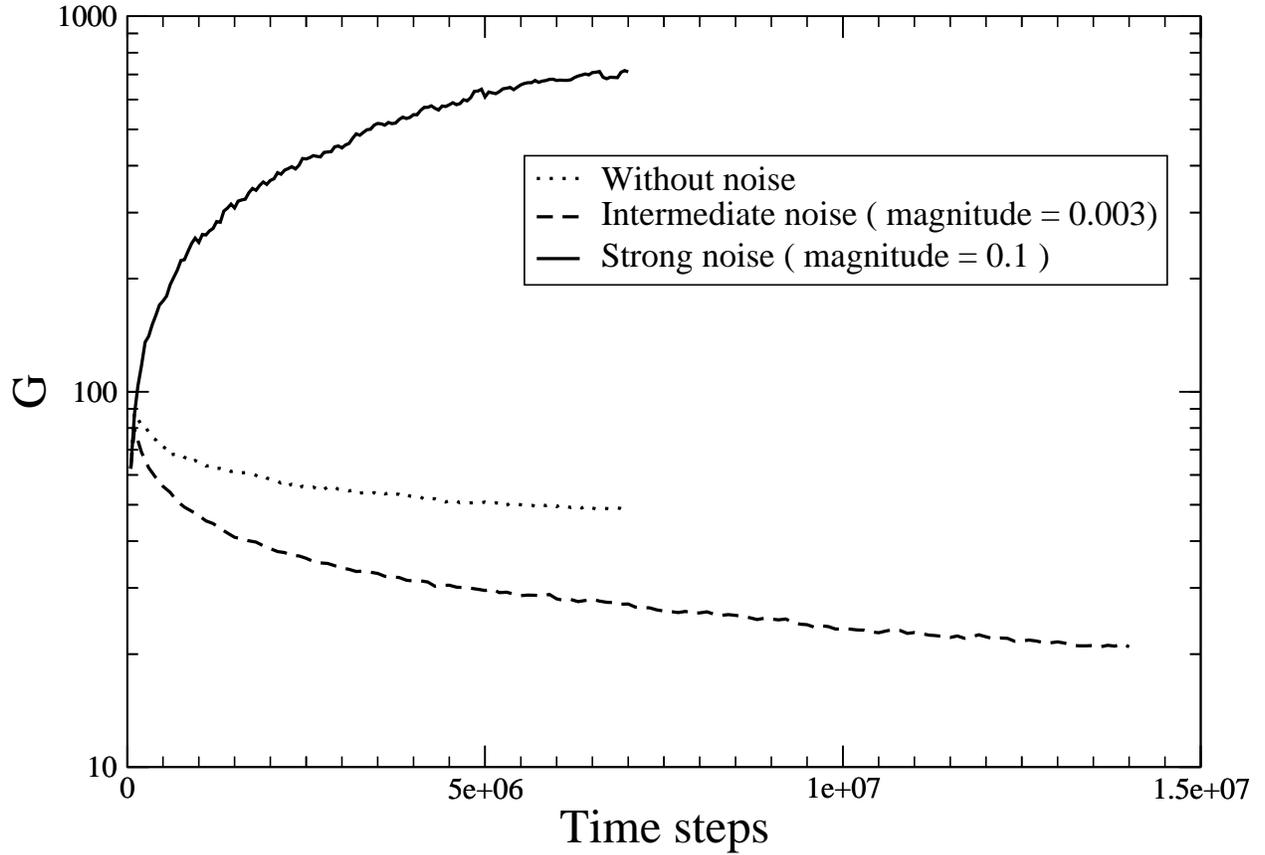}
 \caption{Influence of noise magnitude over model dynamics. Evolution of the number of separated components of the network ($G$), for three
 different values of the noise strength (representative of strong, intermediate and weak noise strength).
 The case without noise is also shown, for comparative purposes. Results have been averaged among 25
 independent realizations.}
 \label{Fig_efecte_soroll}
\end{figure}

In order to answer these questions, we have analyzed the influence
of different noise magnitudes over the quantity of isolated
components forming the network ($G$). In Figure
\ref{Fig_efecte_soroll}, we present the evolution of $G$ for a given
set of initial conditions and different values of the noise
strength. The results corroborate the predicted behaviors for
extremal values of the noise magnitude. Additionally, we observe
that the case corresponding to an intermediate noise strength leads
to steady-states with fewer isolated groups.

Such a surprising result can be related to the capacity of a
moderate social noise to introduce heterogeneity within the
different groups. This internal diversity favors the inter-group
linkage without breaking them into isolated agents. Let's explain
this argument more accurately. When the social noise is weak, the
combined action of imitation and rewiring leads the model to a
steady-state where individuals tend to coincide in a unique $h$
value (social position) and, therefore (because of the rewiring
action), to conform a unique connected component. On the contrary,
when the magnitude of the social noise is extremely high,
differences between $h$ values of agents (social distances among
them) grow such quickly that cannot be counteracted by the imitation
mechanism and, when those distances are too large to maintain links
between neighbors, groups are progressively dissolved towards a
completely disconnected scenario. In an intermediate situation, the
noise intensity is high enough to maintain a wide variety of $h$
values, but the differences introduced among these values are small
enough to keep agents linked
and, in some cases, to establish new links with agents belonging to other groups.\\

\subsection{Experiment}

In order to check the utility of our model as a tool to study the
concept of social cohesion, we have conducted an experiment
comprising two crisis cycles (sudden increases of the social
temperature followed by longer reactionary periods). Each one of the
crisis cycles has consisted on a short period (about 50000 time
steps) of high social temperature (high values of $b$), followed by
a fall to extremely low values of $b$ (reproducing an habitual
reactive behavior of populations after an emergency situation) and,
finally, a progressive recovery towards normality. The $b$ values
chosen to represent each period are 0.5 for 'normal' social
temperature, 2.0 for highly conflictive situations and
0.25 - 0.35 for the reactionary intervals.\\

\begin{figure}[t]
\plotone{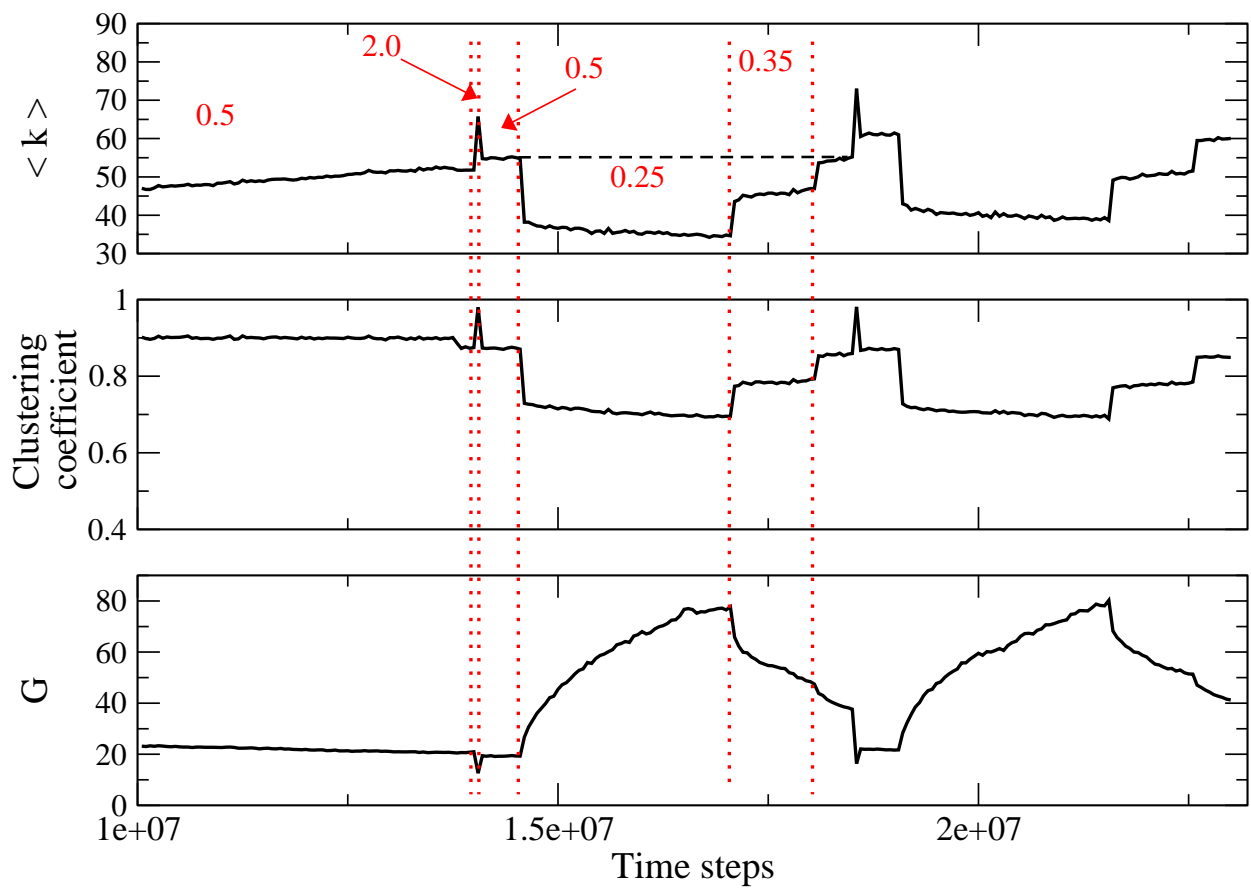}
 \caption{Evolution of social cohesiveness of a population of $N=1000$ agents during the experiment.
 Vertical dashed lines in red indicate regions delimited by their $b$ value, which are indicated also
 in red. The other main parameters of the model, $\alpha$ and the noise magnitude, were set to 6 and 0.003,
 respectively. Two important phenomena are observed: An increase on the average cohesion after each crisis,
 and a memory effect in the period between crises (represented here by a horizontal dashed line in black).
 Results were obtained by averaging 25 independent realizations.}
 \label{Fig_resul}
\end{figure}

When looking at the behavior of the observables during the
experiment, shown in figure \ref{Fig_resul}, we observe two
phenomena. First we notice that, for the same value of $b$, the
social cohesiveness after each emergency situation is higher than
before them. Second, we observe a memory effect on the cohesion of
the system in the period between crises. Although the cohesiveness
diminishes as a response to social temperature cooling, when the
situation comes back to normality, the cohesiveness also recovers
its "normal" value (that one corresponding to $b=0.5$ just after the crisis).\\

The first result agrees with one of the observations made when we
analyzed the three empirical cases, in the sense that the structure
of the system changes during the conflict period. Moreover, it can
be positively contrasted with observations of real social systems.
When a population has been submitted to a stressing situation, it is
quite usual to find higher levels of cohesion than before the
crisis. In some sense, this phenomenon could be seen as a sort of
\emph{reminiscence} of the high rates of cohesion characterizing the
emergency situation.\\

\subsection{Analyzing mesoscopic and microscopic dynamical aspects of social cohesion}
Up to this point, our model has revealed its capacity to reproduce how changes on the social temperature (which is a macroscopical variable related to the social environment) induces changes on the cohesiveness of a population of individuals (here measured in terms of macroscopical observables).\\ 

Nevertheless, in the previous chapter we have argumented that the analysis of the concept of social cohesion from a dynamical viewpoint demands a more complete scope of the problem, including also the behavior of different variables at meso and micro levels during the conflict period. In order to deep in this issue, we have studied how our model's dynamics modifies the distribution of agents' positions ($h_i$ values) along the lineal social
space and, consequently, how the social structure of the population is transformed.\\

\begin{figure}[t]
\plotone{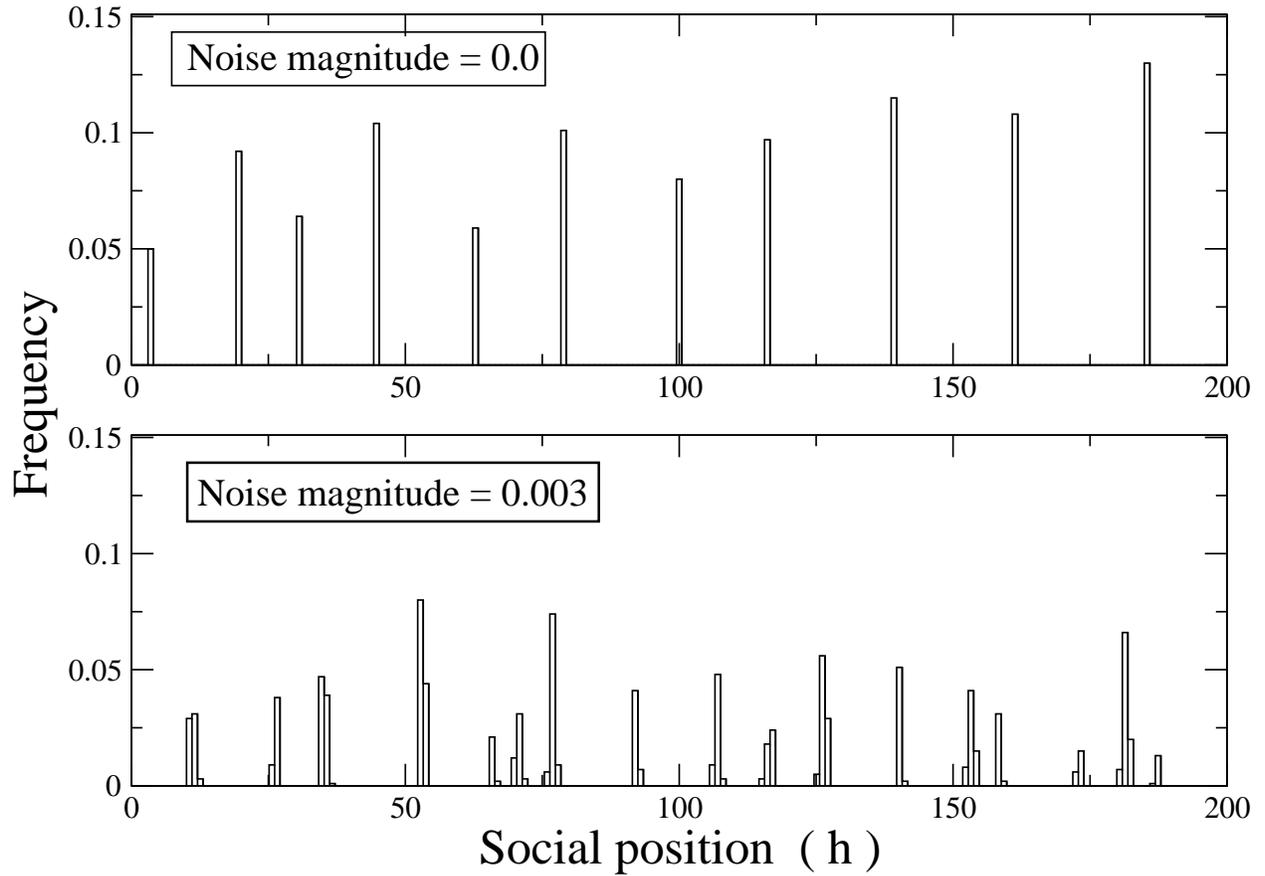}
 \caption{Distribution of agents' positions along the lineal social
space, in a steady-state,  without social noise (top) and with a
social noise of magnitude 0.003. Although both cases present quite
regular separations between groups (see text for an explanation),
the internal distribution of each group differs. In the bottom case,
we appreciate the heterogeneity within groups introduced by the
social noise.} \label{Fig_hs}
\end{figure}

In general, when plotting the distribution of agents' opinions in
the social space at a steady-state (see Fig. \ref{Fig_hs} for two
particular examples), we find that agents' positions are grouped
around certain positions of the space, and that there are quite
regular separations among these concentrations. Taking into account
the dependence of the linkage probability on the social distance, we
deduce that these concentrations of opinions in the social space
correspond, structurally speaking, to groups of agents densely
connected. Besides, the observed separations tend to a unique value
that we have called \emph{critical social distance} $d_c(h_i, h_j)$,
which is the maximum social distance at which a link is possible or,
in other words, is the distance to make the link probability close to 0:

\begin{eqnarray}
d_c(h_i, h_j) = \lim_{r\rightarrow 0} d(h_i,h_j)=\lim_{r\rightarrow
0} b\sqrt[\alpha]{\frac{1}{r(h_i,h_j)}-1}\approx
\frac{b}{\sqrt[\alpha]{r(h_i,h_j)}} \label{dist_c}
\end{eqnarray}

From this definition, it is straightforward that links are
established only among agents separated by a distance smaller than
$d_c(h_i, h_j)$. Moreover, the combined effect of imitation and
rewiring makes that any agent located in a social position shorter
than $d_c(h_i, h_j)$ from any group tend to link to that group, and
that two groups tend to merge if they are near enough from each
other. Consequently, in the steady-state not only the distance among
groups, but also their number and size, is related to the critical
social distance. The larger the $d_c(h_i, h_j)$, the fewer separated
groups and the larger the distance among them.\\

Furthermore, by taking a look to expression (\ref{dist_c}), we
realize that \emph{the critical social distance} depends on $b$.
Since this variable controls the \emph{social temperature} in our
model, we can trace a causal path from variations on the social
temperature to structural and knowledge changes experimented by the
population during a crisis period. With this idea in mind, we can
interpret the behavior of the cohesiveness during the experiment
(shown in fig \ref{Fig_resul}) in terms of reductions and increases
of the critical distance, induced by changes on $b$ (that is, the social temperature).\\

At the beginning of the experiment, before the first crisis, the
critical distance is defined by the original $b$ value (0.5). When
the $b$ value becomes 2.0, the critical distance also increases and,
consequently, all agents come across other ones that were previously
out of their range. Globally, this means that the population tend to
reorganize into fewer but larger groups, whose opinions are
separated each other by greater social distances. However, as this
process is interrupted abruptly (due to the briefness of emergency
situations), some agents are 'surprised' halfway between various
groups. After a short transitory period, a new steady-state is
reached. In this new stable scenario, agents conserve many neighbors
of the period before the crisis and have incorporated new ones due
to those agents bridging different groups after the emergency.
Consequently, the resulting groups are larger than before the crisis
and, because of their high internal connectivity, the average degree
and the clustering coefficient also keep higher. This phenomenon is
what we have previously called \emph{reminiscence} of the crisis over the social cohesiveness.\\

At this point, agents are 'trapped' within their groups, their
opinions are too much different from those of other groups to
establish cross-links. Moreover, in this second stable period, the
social noise plays a central role by opening very little internal
discrepancies between members of the same group, that allow the
creation of new groups when the social temperature gets 'colder'
($b$ drops down to 0.25). Later, as the population recovers its
'normal' social temperature (and, therefore, the $b$ value increases
again), the critical distance grows up and little groups tend to
merge and recover the stable configuration reached just after the
crisis, presenting the second phenomenon pointed above, a memory
effect. Finally, during the second cycle, the system presents the
same behavior than in the first one: A higher cohesiveness than before and a memory effect.\\

\section{CASE STUDY}
Now that the model has been thoroughly detailed, it is possible to
regain one of the examples outlined in the introduction and find
out whether the model is relevant to them or not. In particular, we consider Stark's
work on Hungarian economy, taking into account the amount of data it offers. As said, it analyzes the interactions of
firms and parties across an entire epoch of economic and political
transformation from 1987 to 2006 in a case where market-oriented
enterprises and competitive political parties emerged.\\

Before reviewing the mentioned paper, it is necessary to confront its analysis with a brief description of the political context at the same period of time, in order to fully understand the model's descriptive power, since the aim is to match social unrest during that epoch and economic behavior.\\

Soon after World War II, Hungarian parliament passed a new constitution of Hungary modeled after the 1936 constitution of the Soviet Union. The communist era endowed radical nationalization of the economy based on the Soviet model produced economic stagnation, lower standards of living and a deep malaise. However, the Hungarian Uprising in 1956 put a certain halt to Soviet-style politics and in the late 1960s a mixed economy was introduced in Hungary. Market prices and incentives gradually gained ground, and a partial privatization program was initiated. By the end of the 1980s almost half of economic activity was being generated by private business.\\

After 1989 Hungary's emerging market and parliamentary systems inherited a crisis-ridden economy with an enormous external debt and noncompetitive export sectors. The first free parliamentary election, held in May 1990, was won by center-right and liberal party Democratic Forum (MDF). Under Primer Minister J\'{o}zsef Antall, Hungary began to turn to the world market and restructured its foreign trade.\\

P\'{e}ter Boross succeeded as Prime Minister after Antall died in December 1993. The Antall/Boross coalition governments achieved a reasonably well-functioning parliamentary democracy and laid the foundation for a free-market economy, but the massive worsening of living standards because of the free-market reforms led to a massive loss of support. Agriculture was drastically affected and declined by half. A large portion of the iron, steel, and engineering sectors, especially in northeastern Hungary, collapsed. Unemployment, previously nonexistent, rose to 14 percent in the early 1990s but declined after 1994.\\

By the mid-1990s the economy was again growing, but only moderately. Inflation peaked in 1991 and remained high, at more than 20 percent annually, until the second half of the decade. As a consequence of unavoidable austerity measures that included the elimination of many welfare institutions, most of the population lost its previous security; the number of people living below the subsistence level increased from 10 to about 30 percent of the population between 1988 and 1995.\\

The remarkable point after this historical outline is that, although there were negotiated, peaceful political openings, the democratization process in Hungary was not free of conflict. In fact, soon after 1990's first election, the center-right Hungarian government lost most of its popularity. Within six months of the national election of March 1990 their unpopularity was demonstrated in the overwhelming victory of the opposition parties in the local government elections creating a conflictive situation between local and national governments in many areas. Such discontentment endured the first half of the 1990s.\\

\begin{figure}[t]
\plotone{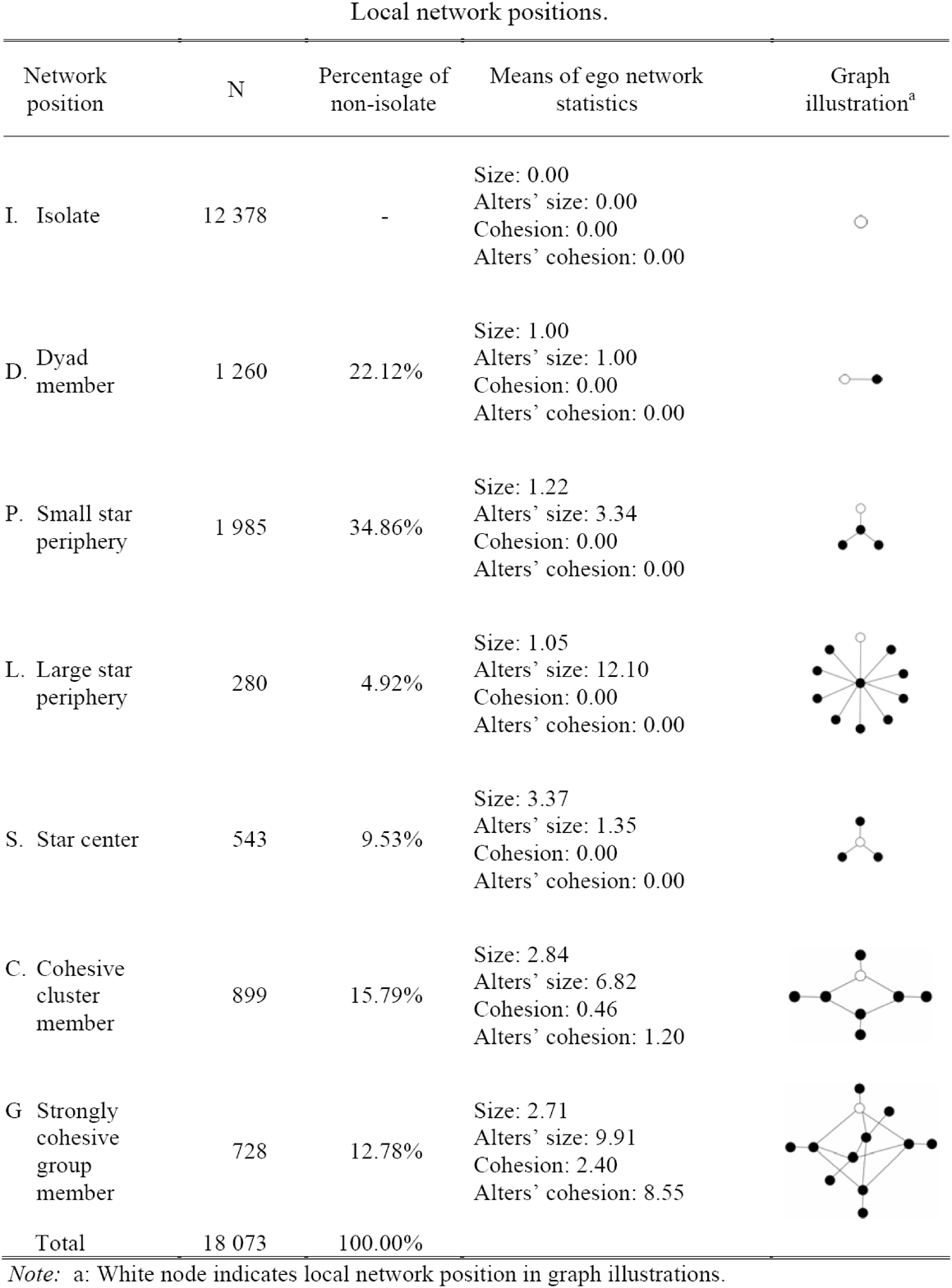}
 \caption{The seven typical local network topographies derived by the cluster analysis: isolate, dyad, small star periphery, large star periphery, star center, cohesive cluster, and strongly cohesive group. \citep{Stark}.} \label{Fig_Stark1}
\end{figure}

\begin{figure}[t]
\plotone{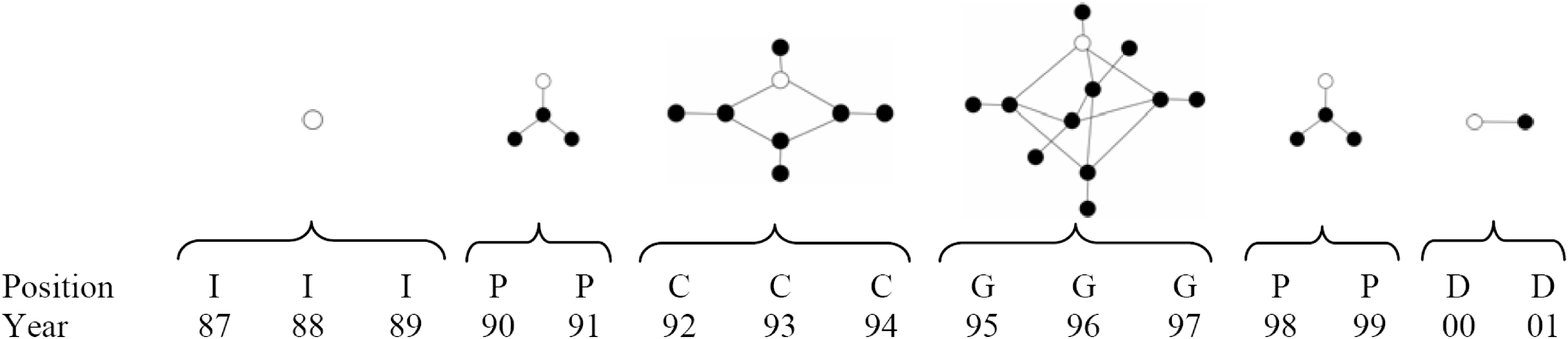}
 \caption{One particular example of firm's history along the period studied. \citep{Stark}.} \label{Fig_Stark2}
\end{figure}

As it has been stated, both theoretically and by means of the simulation model, macro-political variables trigger conflict situations, and communities respond, through information exchange and mobilization at individual level, by increasing structural cohesion. Therefore, the conflictual environment in Hungary during the first years of democratization should match some evidence of cohesion growth. Stark's work on economy is not focused on social movements or protests, but still it presents some useful data that might endorse the point of view here assessed. With attention to temporal sequencing, their work measures cohesion in ownership networks. Fig. \ref{Fig_Stark1} presents the seven typical local network topographies derived by the cluster analysis: isolate, dyad, small star periphery, large star periphery, star center, cohesive cluster, and strongly cohesive group. The particular type of embeddedness for any given firm, in any given year, is now categorized as one of the seven positions. The network history of a firm can now be represented as a sequence of topographies. Fig. \ref{Fig_Stark2} is an example of a firm's history as it moves from one type of embeddedness to another. This firm starts as an isolate. After three years, it becomes the periphery of a small star. In 1992 the topography of the firm's local network is a cohesive cluster, and after three years, these network ties are transformed into a strongly cohesive group. In 1998 the firm becomes a small star periphery again. At the end of the period, from 2000, the star shrinks into a dyad.\\

\begin{figure}[t]
\plotone{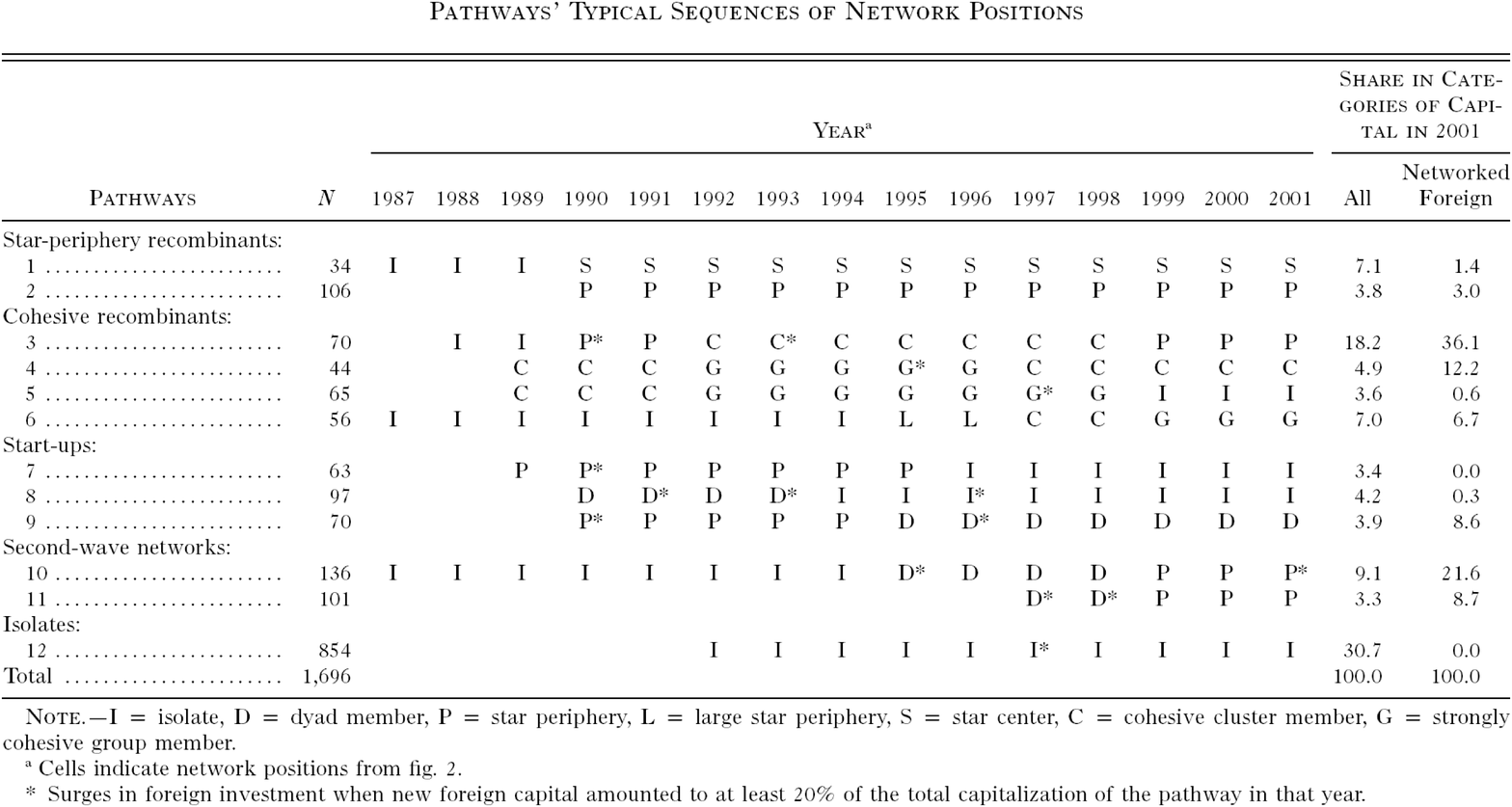}
 \caption{Sequences of network positions that best represents firm histories in each pathway. Reference to Table 2 in the foot-note corresponds to \ref{Fig_Stark2}in this paper. \citep{Stark}.} \label{Fig_Stark3}
\end{figure}

Now that the sequencing methodology is clear, the question whether
social discontent correlates economy processes can be answered.
Stark presents 1,696 such network histories-sequences of positions
for each of the firms in their survey, grouped in terms of their
sequence resemblance (such similar patterns are named
\emph{pathways}). Fig. \ref{Fig_Stark3} presents, for each of the
pathways, the sequence of network positions that best represents
firm histories in that pathway. As it indicates, cohesive
recombinants match our hypothesis about cohesive patterns, and
represent 18,2\% of the firms, i.e. the second largest pathway, only
outnumbered by isolates (one must take into account that small,
family firms are not likely to be involved in ownership changes or
joints). This pattern, and its significance in the whole data
collection, can be interpreted as a delayed reaction to social
non-economic cohesive processes at individual and community level.\\

\section{CONCLUSION AND FUTURE RESEARCH}

In this work we have developed a simple model as an analytic tool to
explore a dynamic perspective of the concept of \emph{social
cohesion}, integrating the already-stated structural component
\citep{Moody&White} with a cognitive, cultural one. Given a certain initial scenario, the model evolves under the influence of the con°ict level of the environment by redefining, simultaneously, the social structure and the knowledge or opinions (represented as positions in a social space) of a population of agents. We argue that, beyond static perspectives, the social cohesion of a population should be expressed in terms of these changes experimented both at structural and cognitive dimensions as a response to conflict increases.\\

By means of only three simple mechanisms, the dynamics of the model
reproduces the behavior of real social populations under a highly
conflictive situation. We have proven this in a threefold way. First
we have studied how changes on a variable of the system representing
the \emph{social temperature} (degree of conflict) conditions the
evolution of three observables than can be easily related to social
cohesion (average degree, clustering coefficient and number of
isolated components). Second, we have deepened in dynamic aspects of
social cohesion by tracing the causal path among different
topological levels: Changes on social temperature happen at an
institutional level, influencing relationships among agents
(microscopical level), and these changes at the individual level
modify the size and composition of groups conforming the social
population (mesoscopic or intermediate level). Finally, third, we
have also compared the quantitative behavior of our model with
empirical observations collected and analyzed in a previous work by \citep{Stark}.\\

Although having demonstrated its utility as a tool to analyze the concept of social cohesion, there are some aspects of the model that could be explored in order to make it more close to particular case studies. In the following, we point out two of these possible extensions of the model.\\

The initial conditions of our experiment, determined by a topology
and a distribution of agents' opinions, can be defined in many
different ways. In this case, we have chosen a simplistic initial
scenario (synthetic social-like topologies and a uniform
distribution of opinions) in order to show that, even starting with
such simple conditions, our model is able to reproduce certain
phenomena related to social cohesion and its dependence on
variations on the social temperature. Nevertheless, each one of the
two components of the initial scenario can be modified separately.
For example, we could use a real social network (obtained by means
of any sort of prospective tool from a real population) as the
initial topology, but we could also start out the experiment with a
distribution of opinions representing a scenario of preexistent
coalitions
or opinion groups.\\

Another possible extension of the model is related to the
observables used to quantify the evolution of the social cohesion.
Although the three structural observables used in this work are too
much simple to represent population's cohesion separately, analyzing
the evolution of their behaviors jointly has helped us to understand
the dynamical processes taking place in the model. Nevertheless, for
the sake of simpleness and clarity, it would be interesting to
define a unique (necessarily more complex) structural observable,
based on previous studies like \citep{White&Harary} and
\citep{Moody&White}. Furthermore, in accordance with the aim of this
work of enriching the structural approach to social cohesion with a
cultural component, it would also be interesting to define an
observable related to the distribution of opinions in the social
space (based on the largest social distance in the system, for instance).\\

\section{ACKNOWLEDGEMENT}
The authors would like to thank

\clearpage

\begin{thebibliography}{}

\bibitem[Blau 1977]{Blau} Blau, Peter M. 1977. \emph{Inequality and Heterogeneity}. New York: The Free Press of Macmillan Co.

\bibitem[Blau \& Schwartz 1984]{Blau2} Blau, Peter M., and Joseph E. Schwartz. 1984. \emph{Crosscutting Social Circles}. Orlando, FL: Academic Press

\bibitem[Boguna et al. 2004]{Boguna&altri} Boguna, Marian, Romualdo Pastor-Satorras, Albert Diaz-Guilera, and Alex Arenas. 2004. " Models of social networks based on social distance attachment". \emph{Phys. Rev. E}, 70 , 056122.

\bibitem[Carley 1991]{Carley} Carley, Kathleen M. 1991. "A theory of group stability". \emph{Am. Sociol. Rev.} 56:331–54

\bibitem[Coleman 1990]{Coleman} Coleman, James. 1990. \emph{Foundations of Social Theory}. Cambridge, MA: Harvard University Press.

\bibitem[Dimaggio 1992]{Dimaggio} DiMaggio, Paul. 1992. "Nadel's paradox revisited: relational and cultural aspects of organizational structure". In \emph{Networks and Organizations: Structure, Form, and Action}, edited by N. Nohria, and R.G. Eccles. Boston: Harvard Business School Press, pp.118-42.

\bibitem[Durkheim 1956]{Durkheim} Durkheim, Emile. 1956. \emph{The Division of Labor in Society}. The Free Press, New York.

\bibitem[Fararo \& Skvoretz 1987]{Fararo} Fararo, Thomas J., and John Skvoretz. 1987. "Unification Research Programs: Integrating Two Structural Theories.", \emph{American Journal of Sociology} 92 (1): 183- 1209.

\bibitem[Forrest \& Kearns 2001]{Forrest} Forrest, Ray, and Ade Kearns. 2001. "Social cohesion, social capital and the neighbourhood". \emph{Urban Studies} 38: 2125–43.

\bibitem[Friedkin 2004]{Friedkin} Friedkin, Noah E. 2004.\emph{Annual Review of Sociology} 30: 409-425.

\bibitem[Gould 1991]{Gould} Gould, Roger V. 1991. "Multiple Networks and Mobilization in the Paris Commune, 1871". \emph{American Sociological Review} 56 (7): 16-29.

\bibitem [Gould 1993]{Gould_paper} Gould, Roger V. 1993. "Collective Action and Network Structure". \emph{American Sociological Review} 58 (2): 182-196.

\bibitem [Gould 1995]{Gould_llibre} Gould, Roger V. 1995. \emph{Insurgent Identities: Class, Community, and Protest in Paris from 1848 to the Commune}. Chicago: University of Chicago Press.

\bibitem[Granovetter 1973]{Granovetter3} Granovetter, Mark. 1973. "The Strength of Weak Ties." \emph{American Journal of Sociology} 68: 1360-80.

\bibitem[Granovetter 1985]{Granovetter} Granovetter, Mark. 1985. "Economic Action and Social Structure: The problem of Embeddedness". \emph{American Journal of Sociology}, 91(3):481-510.

\bibitem[Granovetter 1992]{Granovetter2} Granovetter, Mark. 1992. "Problems of Explanation in Economic Sociology". In \emph{Networks and Organizations: Structure, Form, Action.}, edited by N. Nohria and R. Eccles. Boston: Harvard Business School Press.

\bibitem[Hirsch 1990]{Hirsch} Hirsch, Eric L. 1990. "Sacrifice for the Cause: Group Processes, Recruitment, and Commitment in a Student Social Movement".
\emph{American Sociological Review} 55 (2): 243-254.

\bibitem[Holme \& Newman 2006]{Holme&Newman} Holme, Peter, and Mark E.J Newman. 2006. "Nonequilibrium phase transition in the coevolution of networks and opinions". \emph{Phys. Rev. E} 74, 056108.

\bibitem[Kawachi \& Kennedy 1997]{Kawachi} Kawachi, Ichiro, and Bruce P. Kennedy. 1997. "Health and social cohesion: why care about income inequality?" \emph{British Medical Journal} 314: 1037-1040.

\bibitem[Kearns \& Forrest 2000]{Kearns} Kearns, Ade, and Ray Forrest. 2000. "Social Cohesion and Multilevel Urban Governance". \emph{Urban Studies} 37 (5–6): 995–1017.

\bibitem[Lazer 2001]{Lazer} Lazer, David. 2001. "The co-evolution of individual and network". \emph{J. Math. Sociol} 25:69–108

\bibitem[Marsili et al. 2004]{Vega-Redondo} Marsili, Matteo, Fernando Vega-Redondo, and Frantisek Slanina. 2004. "The rise and fall of a networked society: A formal model". \emph{Proc. Nat. Acad. Sci.} 101: 1439-1442.


\bibitem[McPherson \emph{et al.} 2001]{McPherson} McPherson, Miller, Lynn Smith-Lovin, and James M. Cook. 2001. "Birds of a feather: Homophily in Social Networks". \emph{Annu. Rev. Sociol.} 27: 415-44.

\bibitem[Moody \& White 2003]{Moody&White} Moody, James, and Douglas R. White. 2003. "Structural Cohesion and Embeddedness: A Hierarchical Concept of Social Groups". \emph{American Sociological Review} 68 (1): 103-127.

\bibitem[Murphy 1957]{Murphy} Murphy, Robert F. 1957. "Intergroup Hostility and Social Cohesion". \emph{American Anthropologist} 59 (6): 1018-1035.

\bibitem[Nadel 1957]{Nadel} Nadel, S.F. 1957. \emph{Theory of Social Structure}. London: Cohen and West. 

\bibitem[Polanyi 1944]{Polanyi} Polanyi, Karl. 1944. \emph{The Great Transformation}. New York: Rinehart and Co.

\bibitem[Putnam 1993] {Putnam} Putnam, Robert. 1993. \emph{Making Democracy Work. Civic Traditions in Modern Italy}. Princeton: Princeton University Press.

\bibitem[Putnam 1995]{Putnam2} Putnam, Robert. 1995. "Bowling Alone: America's Declining Social Capital". \emph{Journal of Democracy} 6 (1): 65-78.

\bibitem[Putnam 2001]{Putnam3} Putnam, Robert. 2001. "Social Capital: Measurement and Consequences". \emph{Isuma-Canadian Journal of Policy Research} 2 (1): 41-51.

\bibitem[Radcliffe-Brown 1940] {Radcliffe-Brown} Radcliffe-Brown, A.R. 1940. "On Social Structure". In \emph{Social Networks; a Developing Paradigm}, edited by Samuel Leinhardt in 1977. New York: Academic Press. pp. 221-232.

\bibitem[Room 1995]{Room} Room, Graham (ed.). 1995. \emph{Beyond the Threshold: The Measurement and Analysis of Social Exclusion Bristol}. The Policy Press.

\bibitem[Snijders 2005] {Snijders} Snijders, Tom A.B. 2005. "Models for Longitudinal Network Data". In \emph{Models and methods in social network analysis}, edited by P. Carrington, J. Scott, and S. Wasserman. New York: Cambridge University Press.

\bibitem[Snijders et al. 2006] {Snijders&altri}. Snijders, Tom A.B.; Christian E.G Steglich, and Michael Pearson. 2006. \emph{Dynamic Networks and Behavior: Separating Selection from Influence}. Groningen: RuG.2007. 

\bibitem[Snow et al. 1986]{Snow} Snow, David A., E. Burke Rochford, S.K. Worden, and Robert D. Benford. 1986. "Frame Alignment Processes, Micromobilization, and Movement Participation". \emph{American Sociological Review} 51 (4): 464-481.

\bibitem[Stark \& Vedres 2006]{Stark} David Stark and Barázs Vedres. 2006. "Social Times of Network Spaces: Network Sequences and Foreign Investment in Hungary". \emph{American Journal of Sociology} 111 (5): 1367–1411

\bibitem[Wasserman \& Faust 1994]{Wasserman} Wasserman, Stanley, and Katherine Faust. 1994. \emph{Social Network Analysis: Methods and Applications}. Cambridge: Cambridge University.

\bibitem[Wasserman \& Robins 2005] {Wasserman&Robins} Wasserman, S., and  G.L. Robins. 2005. "An Introduction to Random Graphs, Dependence Graphs, and p*". In  \emph{Models and Methods in Social Network Analysis} (pp. 148-161), edited by P. Carrington, J. Scott, and S. Wasserman. New York: Cambridge University Press

\bibitem[Wellman \& Leighton 1979]{Wellman&Leighton} Wellman, Barry, and Barry Leighton. 1979. \emph{Urban Affairs Quarterly}, 14(3): 363-390.

\bibitem[Wellman et al. 1996]{Wellman&altri} Wellman, Barry, Janet W. Salaff,  Dimitrina Dimitrova, Laura Garton, Milena Gulia, and Caroline Haythornthwaite. 1996. "Computer Networks as Sociel Networks: Collaborative work, Telework and Virtual Community". \emph{Annual Review of Sociology} 22: 213-238.

\bibitem[White \& Harary 2001]{White&Harary} White, Douglas R., and Frank Harary. 2001. "The cohesiveness of blocks in Social Networks: Node connectivity and Conditional Density". \emph{Sociological Methodology}, 31: 305-359.

\bibitem[White and al. 2004]{White&altri} White, Douglas R., Jason Owen-Smith, James Moody, and Walter W. Powel. 2004. "Networks, Fields and Organizations: Micro-Dynamics, Scale and Cohesive Embeddings". \emph{Computational
and Mathematical Organization Theory}, 10 (1): 95-117.

\bibitem[Wilkinson 1996]{Wilkinson} Wilkinson, Richard G. 1996. \emph{Unhealthy Societies: The Afflictions of Inequality}. London: Routledge.

\bibitem[Wolfe 2002]{Wolfe} Wolfe, David A. 2002. "Social Capital and Cluster Development in Learning Regions" in \emph{Knowledge, Clusters and Learning Regions}, edited by J. Adam Holbrook and David A. Wolfe. Kingston: School of Policy Studies, Queen's University.

\end{thebibliography}
\end{document}